\documentclass{article}

\usepackage{fullpage,amsmath,epsfig}

\title{DyNoC: A Dynamic Infrastructure for \\
Communication in Dynamically Reconfigurable Devices
  \thanks{Supported in part by the German Research Foundation (DFG), SPP 1148 (Rekonfigurierbarer Rechensysteme)}}

\author{
  Christophe~Bobda\thanks{
  Department of Computer Science 12,
  University of Erlangen-Nuremberg,
  Germany.
 \{bobda, ahmadinia, majer, teich\}@cs.fau.de.}
 \and Ali~Ahmadinia\footnotemark[2]
  \and Mateusz~Majer\footnotemark[2] \and J\"urgen Teich\footnotemark[2]
  \and S\'andor~Fekete\thanks{
 Institute of Mathematical Optimization,
Braunschweig University of Technology, 
Germany.
\{s.fekete,j.van-der-veen\}@tu-bs.de.}
\and  Jan~van~der~Veen\footnotemark[3]}
\begin{document}

\maketitle

\begin{abstract}
A new paradigm to support the communication among modules dynamically placed on a reconfigurable device at run-time is presented. Based on the network on chip (NoC) infrastructure, we developed a dynamic communication infrastructure as well as routing methodologies capable to handle routing in a NoC with obstacles created by dynamically placed components. We prove the unrestricted reachability of components and pins, the deadlock-freeness and we finally show the feasibility of our approach by means on real life example applications.  
\end{abstract}
\section{Introduction}
On-line placement methods on reconfigurable device like that described in \cite{ABFTV04} have been around for a while. 
On-line placement is interesting as long as all incoming components are free 
entities without connection to the rest of the world, i.e. the component does not 
communicate with other components, an unrealistic situation. Imagine a new component 
placed at run-time in the middle of the chip, completely surrounded by other 
components and expecting its data from some pins around the device. 
It is therefore mandatory to develop methods to allow components placed 
at run-time on the device not only to communicate with others, but also 
to have access to the pins needed for a data exchange with off-chip devices.
In this paper, we address the problem which arises when components dynamically
placed on a reconfigurable device need to communicate with other components
on the chip or with off-chip modules. Few authors \cite{ABFTV04} have recently addressed this 
problem. However, they focus on developing a solution for the Xilinx FPGA which
provides only a 1-D temporal placement model. We present a new NoC-based architecture 
which allows an unlimited communication between components and pins. We further present 
a new routing approach based on the well known XY-routing and modify it to handle obstacles 
in the network. We prove the correctness of our method and show that each component as 
well as each pin is reachable from other components. Finally, we present the result of
implementation on real life applications.\\
We organize the paper as follows: In Section \ref{relwork}, we present related work on
communication on dynamic on-chip network. Section \ref{dynoc-basics} introduces the 
DyNoC architecture as well as the dynamic connection of components in the
network. In section \ref{xyrouting}, we propose an extension of the XY-routing
able to handle obstacles in the network and we prove its feasibility. 
Section \ref{casestud} handles a case study. There real life applications are implemented 
in the FPGA-board 
RC200 of Celoxica using the DyNoC. Finally, section \ref{conclusion} summarizes the paper 
and provides some indications on future work.
\section{Related work}\label{relwork}
Run-time communication support on dynamic reconfigurable devices like FPGAs has
been recently addressed in several papers.
Some approaches like \cite{TM03b} assume a partitioning of the device in logical blocks or 
bins in which the incoming modules must be placed. 
The communication link is set at compiled time and no dynamics is used in the model.
The authors have focused on the development of solutions for the Xilinx 1-D. 
Their solutions cannot be extended to a 2-D model. 
In order to be able to establish communication among components dynamically
placed at run-time on a device, the device itself must provide a viable 
communication infrastructure. Two possibilities exist here: The first one is 
the so called circuit routing which allows two modules which are willing to 
communicate to establish a physical connection by setting some switches on the 
communication links. This approach presents some major drawbacks: First, computing 
a route at run-time is expensive. Secondly, already established connections 
limit the routing possibilities for a new route and therefore jeopardize setting the run-time 
connection. The second possibility is the network on chip paradigm, which allows two modules to communicate by sending packets instead of a direct connection. 

In a network on chip, several modules (network clients) placed at fixed locations 
can exchange packets in the common network. This provides a very high flexibility, 
since no route has to be computed before allowing components to start communicating. 
Components just send packets and they don't care on how the packets are routed in 
the network. Networking on chip is viewed as the ultimate solution to avoid problems which will arise due to the growing size  of the chip. 
Networking on a chip presents a viable communication
infrastructure, however it is still too inflexible to dynamically support the 
communication among modules in a changing network. Each module must be placed on one of the bin, i.e., implemented on one processing element (PE), connected to a
network element (also known as router) for accessing the network. Large
modules, which cannot fit on one PE must be implemented on a set of neighbor
PEs. The communication among the different parts of the module will therefore be
packet-based, thus increasing the complexity of the module and wasting more
resources. 
A better implementation would connect all the PE using
direct wiring. This has no disadvantage, because the PEs are close to each
other and therefore long connections are avoided.  
In our concept, the routers inside the boundary of a module are redundant. 
They can be used as additional resources to implement an even bigger module. 
This can be achieved if the routers are programmable elements which are set 
in their basic configuration to behave as routers. 
The placed module will then access the network using only one router. 
With this, the NoC becomes dynamic, thus
allowing modules to be placed on large area and use the underlying routers
logic. This concept, called DyNoC (Dynamic Network on Chip) was first
presented in \cite{BMKAT04}. However, the routing strategies were not
investigated and no evaluation of the concept with real life problems was
done.\\
Routing in networks is a very old topic and a well understood research area. Obviously, 
some work has been done in routing packets in a dynamically changing network.
Existing work rely on learning-algorithms like Q-learning \cite{boyan94packet}, 
a special case of reinforcement learning. With reinforcement learning in general 
and Q-learning in particular, each router is an autonomous structure which learns 
with the time the most efficient route to all possible destinations.  
With a frequently changeable network, the router will spend most of its time for 
learning the new network structure, thus 
decreasing the network performance. Furthermore, the complexity of those algorithms 
does not qualify them to be used on a chip. The well known greedy XY-algorithm usually
performs well in practice and routes packets according to the Manhattan distance.
However, no existing work has addressed the extension of the XY-routing
for dealing with a) changing networks and b) being different from a grid. In this work, we
present a new routing approach based on the well known XY-routing. We prove
the correctness of our method and show that each component as well as each pin
is reachable from other components. Our method can be used for all other
network topologies. The feasibility of our approach is tested on real life
applications.
\section{The Dynamic Network on Chip ({DyNoC})} \label{dynoc-basics}
 Many well reputated authors \cite{benini,hemani00network} have
 predicted that wiring modules on chip will not be a viable solution in the billion
 transistor chips in the future. Instead, they proposed Networks-on-Chip (NoC)
 as a good solution to support communication on System-on-Chip in the
 future. NoCs encounter many advantages (performance, structure and
 modularity) toward global signal wiring. A chip employing a NoC is composed of
 a set of network clients like DSP, memory, peripheral controller, custom
 logic which communicate on a packet basis. However, fixed NoCs are not flexible enough for supporting communication in a dynamically changing 
 network on chip. We present in the next section the main modifications we considered 
 on a NoC.
\subsection{Communication infrastructure} \label{archrequire}
The goal is to have a communication infrastructure in which the reachability
of packets is ensured, independent of the changing topology which occurs when
components are placed and removed on the chip. In its basic state, the
communication infrastructure is a normal NoC. Processing elements
access the network via a network element. Additionally, direct communication paths
exist between neighbor PEs. In this way, the network elements are only used
for communication between non-neighbor PEs. As stated earlier, the placement of 
a module in a given region of the chip makes the routers in that region useless, 
since PEs belonging to the module are directly connected. The idea is then to 
implement routers as reusable elements which behave as routers in their basic
configuration, but can be used by a component as part of its logic. Such router
can be available as programmable hard macro on the chip. Whenever a component
is placed in a given region, only one router is necessary for this element to access 
the network. Without loss of generality the router attached to the upper 
right PE of the module is used. 
\subsection{Network access}\label{Netaccess}
Each task is implemented as a component, represented by a rectangular box 
and stored in a database. Since synthesis is a time consuming task, it cannot
be done on-line. Therefore, the synthesis of components is done at compile
time. A box encapsulates a circuit implemented with the resources in a given
area (routers logic and PEs). After the placement of a new component on the device,
its coordinate is set to that of its corresponding router. When placed on the device,
components hide part of the network which is restored when they
complete their execution. This makes the network dynamic. This is why we call
such a network a  {\bf {\em dynamic network-on-chip (DyNoC)}}.
\subsubsection{Reachability of components and pins}\label{netaccess}
\begin{figure}[t]
\begin{minipage}[b]{1.0\linewidth}\centering
 \centerline{\epsfig{figure=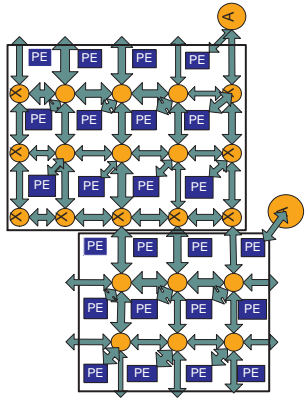,width=40mm}}
\end{minipage}
\caption{Non valid placement on the DyNoC}\label{wrong_placed}
\end{figure}

\begin{figure}[t]
\begin{minipage}[b]{1.0\linewidth}\centering
 \centerline{\epsfig{figure=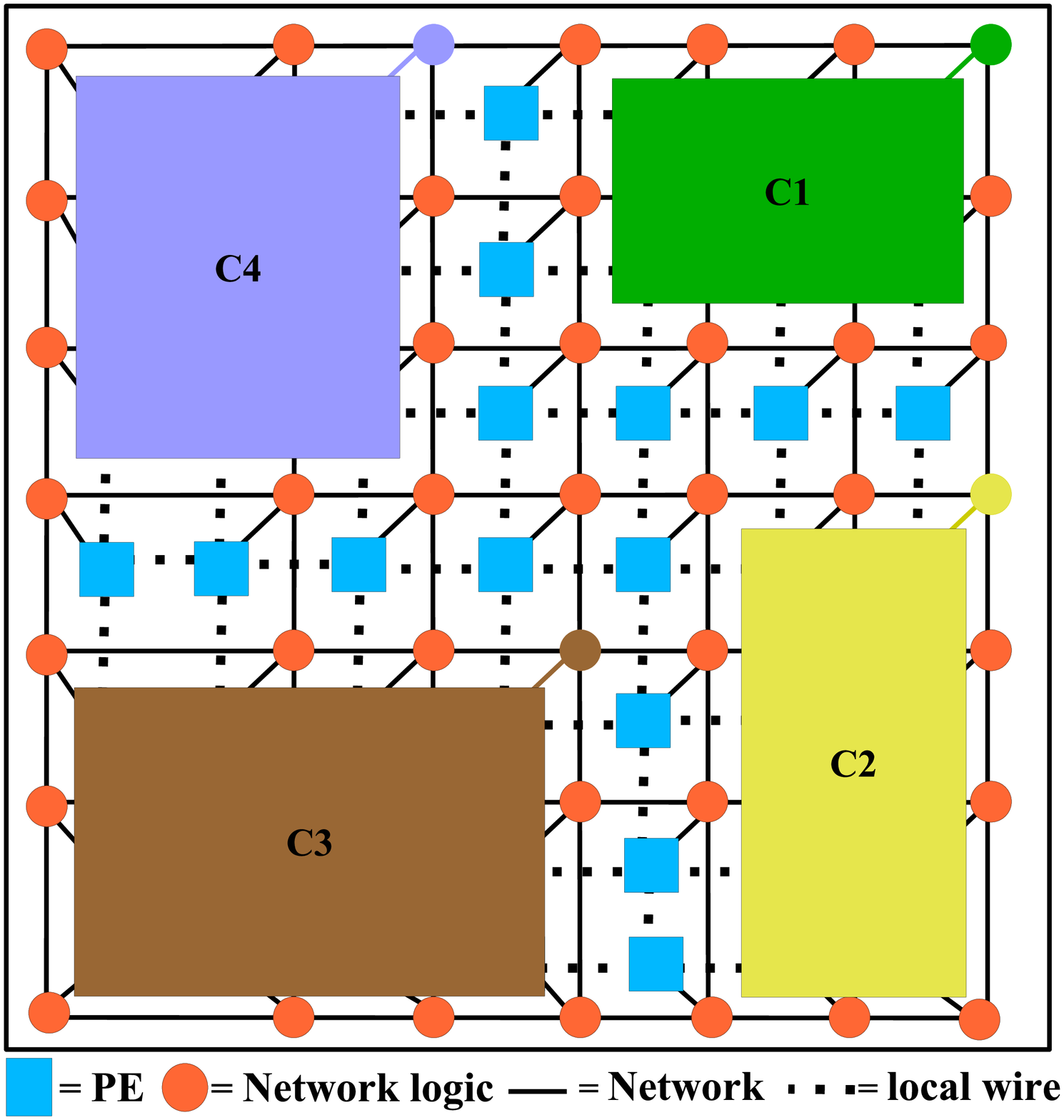,width=40mm}}
\end{minipage}
\caption{Valid placement on the DyNoC}\label{dynoc_placed}
\end{figure}

We said that a component (pin) on reconfigurable device at a given time is
reachable iff each message sent to this component (pin) can reach the
component (pin). Because the communications between components are established at 
run-time and since the configuration\footnote{We define the configuration of the device 
as the set of components actually running on the device} of the chip is not known in 
advance we must insure that all components and pins on the device are reachable at any time 
during the temporal placement. This condition is fulfilled if at any time the set of 
components and pins on the device is strongly connected\footnote{A set of
components is said to be strongly connected, iff for each pair of components a path of routers
exists which connects the two components}. One way to enforce this is to require
that each component placed on the chip must always be surrounded by a ring of 
routers. This can be reached either by synthesizing components in such a way
that when placed on the device, they are always surrounded by a ring of routers.
The second way is to let the job do by a temporal placer. This will considerably
increase the complexity of the placer. Besides the computation of free space
to place a new component it must be ensured that the placement is strongly
connected. We therefore opt for the first solution.
\newtheorem{theorem}{Theorem}
\begin{theorem}
If each component is synthesized in such a way that it is internally surrounded only
by processing elements, then each placement on the reconfigurable device is strongly 
connected.
\label{theo1}
\end{theorem}
\textbf{Proof:}
\textit{
Assume that a set of components developed as require in Theorem \ref{theo1} 
and placed on the device is not strongly connected. In this case, a) at least one 
pair of components abuts or b) a component abuts the device boundary. Let's consider 
the first case. The second one can be handled in a similar way. Either the two 
components overlap or at least one component use some routers on its internal
boundary (this is illustrated in Figure \ref{wrong_placed}) . The first case
is impossible because only overlapping free placements are valid. The second
case contradicts our requirement of the theorem, thus completing the proof.\\
}
Figure \ref{wrong_placed} illustrates an impossible placement scenario where
two components abut while Figure \ref{dynoc_placed} shows a placement in
which all components and pins are reachable.

While in a static NoC, each router always has four active neighbor routers\footnote{The neighbor routers of the routers around the chips are assumed to 
be the package pins through which external modules can access the network},
this is not always the case in the DyNoC presented here. Whenever a component 
is placed on the device, it covers the routers in its area. Since those routers 
cannot be used, they are deactivated. The component therefore sets a
(de)activation signal to the neighbor routers to notify them not to send
packets in its direction. Upon completing its execution, the deactivated
routers are set to their default state. A routing algorithm used for common
NoC cannot work on the DyNoC. We need therefore either to modify existing
algorithms or develop new ones.
\section{Routing Packets}\label{xyrouting}
In the DyNoC, we face a new situation. With the dynamic placement and removal 
of modules on the chip, unpredictable obstacles are created. The routing
algorithm must be able to deal with this situation. The router algorithm must
be fully local-decisive\footnote{The decision where to send a packet is taken at
the local level} and deadlock-free\footnote{Each packet will reach its destination 
after a finite number of steps}.\\  
Due to its simplicity, its efficiency and its deadlock-freeness, we have chosen to 
adapt the XY-routing algorithm for the DyNoC. In a full mesh, XY-routing is a 
deadlock-free shortest path routing algorithm that first routes packets in X direction 
to the correct X-coordinate and then in the Y direction until the correct Y-coordinate. 
In the DyNoC, the placement of components at run-time alters some parts of the grid, 
thus producing {\em obstacles} into the mesh. We adapted the XY-routing to deal with
obstacles. In our new algorithm called ``S-XY-Routing'' (Surrounding XY routing)
the routers operate in three different modes: The first mode is the N-XY (Normal XY) mode. In this mode, the router behaves as a normal XY-router. The second is the SH-XY (Surround horizontal XY) mode. The router enters this mode when its left neighbor or its right neighbor is deactivated. The third mode is the SV-XY (surround vertical XY) mode. The router enter this mode when its upper neighbor or its lower neighbor is deactivated. In the N-XY mode, the packets are first sent horizontally to their right X-coordinates and then vertically routed to their Y-coordinates. 
As we will see later, horizontal obstacles should be treated differently than vertical ones.
\subsection{Surrounding Obstacles in the X-direction}
\begin{figure}[t]
\begin{minipage}[b]{1.0\linewidth}\centering
 \centerline{\epsfig{figure=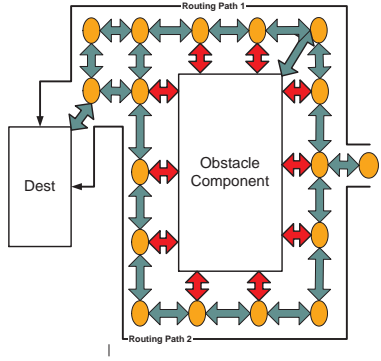,width=50mm}}
\end{minipage}
\caption{Obstacle avoidance in the horizontal direction}\label{dynoc-obstacle-x}
\end{figure}

Assume without loss of generality that a packet moving from right to left is
blocked by an obstacle. As shown in Figure \ref{dynoc-obstacle-x}, there exist
two alternative paths for the packet to reach its destination according to
the Y-coordinate. 
The first path is chosen if the Y-coordinate of the destination of the packet is 
greater or equal than that of the router and the packet is sent upwards. Otherwise, 
the second path is chosen and the packet is sent downwards. One problem occurs when
a packet with destination $Y_{dest}$ is sent for example upwards and reaches a router 
$r$ with coordinate $Y_{r} > Y_{dest}$. According to the previously defined scheme
the packet will be sent downwards to the router with coordinate  $Y_{r}-1$ which 
will send it upwards, thus producing a "`ping-pong" effect. To avoid this, we stamp the 
packet by setting a "stamp-bit" to 1 to notify router $r$ not to send the packet back. 
Upon reaching the router upper right to the device, the stamp is removed 
and the packet is sent left, until its destination column or until another obstacle is 
found. In the example of Figure \ref{dynoc-obstacle-x}, path 1 will be chosen.\\
Because each component is always surrounded by a ring of routers, our
algorithm for surrounding a component in the X-direction will always work and
a packet will never be blocked.

\begin{figure}[t]
\begin{minipage}[b]{1.0\linewidth}\centering
 \centerline{\epsfig{figure=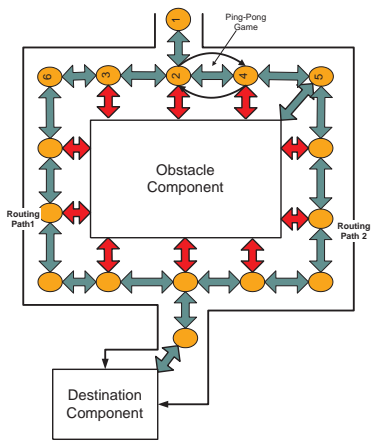,width=50mm}}
\end{minipage}
\caption{Obstacle avoidance in the verticalal direction}\label{dynoc-obstacle-y}
\end{figure}
\subsection{Surrounding Obstacles in the Y-direction}
\begin{figure}[t]
\begin{minipage}[b]{1.0\linewidth}\centering
 \centerline{\epsfig{figure=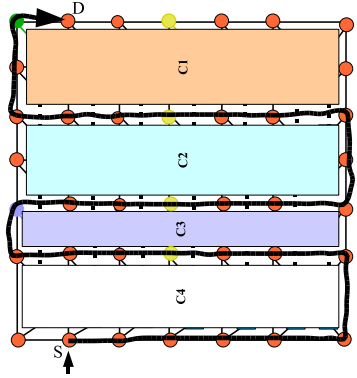,width=50mm}}
\end{minipage}
\caption{Placement situation with extreme long routing path}\label{dynoc-extrem}
\end{figure}

The situation is different when a packet moving in the Y-direction is
blocked. Assume without loss of generality that a packet moving from top to 
bottom is blocked by a placed component. Dealing with this case as the
previous one, the packet will be sent left or right. No preference is imposed 
here, because the packet is already in its right column. Lets assume that 
the packet is sent to the right to the next router. Because the basic routing 
algorithm is the XY-routing, the next router will first compare the X position 
of the packet with its own position. With the packets X-destination being
smaller, it will send the packet back to the router from which it received the 
packet. The two routers will keep sending the same packet to each other, thus 
creating a deadlock. Figure \ref{dynoc-obstacle-y} in which a ``ping-pong'' 
reflect results between router 2 and router 4 illustrates this situation. 
To avoid this ``ping-pong'' game, we stamp the packet to notify all the routers
 above the obstacle that the packet is willing to surround the component. In our 
 example, the packet will then be sent right until the last router (5) above the 
 component. There, the router removes the stamp and sends the packet downwards. 
 From there on, we have the same situation as defined in the previous section 
 (Surrounding Obstacles in the X-direction). Since we could show that the packet 
 will always reach its destination in the previous section, we conclude that the 
 packet will also reach its destination in this case.
\begin{theorem}
With a very high probability, the S-XY algorithm presented here is deadlock free.
\label{theo}
\end{theorem}
\textbf{Proof:}
\textit{
We need to first prove that there is always a path from the source of a packet to its destination.
Second we must prove that each packet will reach its destination after a fixed
number of steps, if no component is placed in between. The first requirement is 
guaranteed through Theorem \ref{theo1}.
We now assume that a packet never reaches its destination, if no component is placed 
in between. This will happen only if the packet is blocked or if the packet is looping in a given region. Because a 
path always exists from one active router to all other active routers, no packet 
can be blocked in the network, i.e. a packet is looping. Since this situation is not 
possible in the normal XY, it can only arise in the surrounding phase. 
When a packet is blocked in a given direction, it takes the perpendicular direction. 
This is done until the last router on the component boundary which is at one corner 
of the module to be surrounded. From there, the normal XY routing resumes. Looping 
of a packet around a component is therefore not possible.
Obviously, a placement with a set components can be constructed in such a way that a 
packet keeps going around in the device. However, all the remaining packets will 
behave as if nothing has happened. In such a case, only one packet over all the other
will be lost. The probability for a packet to be blocked infinitely is then very low.\\ 
}
In the S-XY routing, fixing a priori for all routers the direction where to send a 
packet whenever an obstacle is encountered can lead to extreme long routing paths like that of Figure \ref{dynoc-extrem}, caused by placements for which the routers 
always choose the extreme longest path.
\begin{figure}[t]
\begin{minipage}[b]{1.0\linewidth}\centering
 \centerline{\epsfig{figure=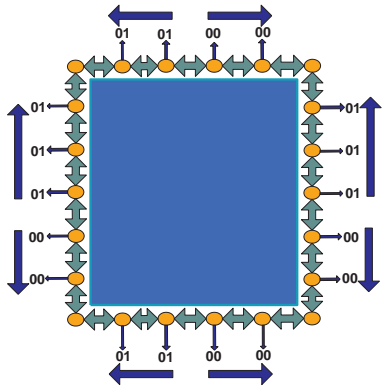,width=50mm}}
\end{minipage}
\caption{Router guiding in a DyNoC}\label{dynoc-guide}
\end{figure}
To avoid this, each router is instructed by the placed component about the direction 
to take whenever an incoming packet is blocked in a given direction by the component. 
Instead of using only one activation line, two lines are used in this case. Figure \ref{dynoc-guide} illustrate this approach. There, the first line is used for the activation (1=activate, 0=deactivated) and the second one for the direction to take (0=(east or south), 1=(west or north)). This considerably limits the complexity of the routers and there is no need for stamping anymore. We call this modification  \textit{router guiding} because the routers are guided by the components.
\section{Case study}\label{casestud}
\begin{figure}[t]
\begin{minipage}[b]{1.0\linewidth}\centering
 \centerline{\epsfig{figure=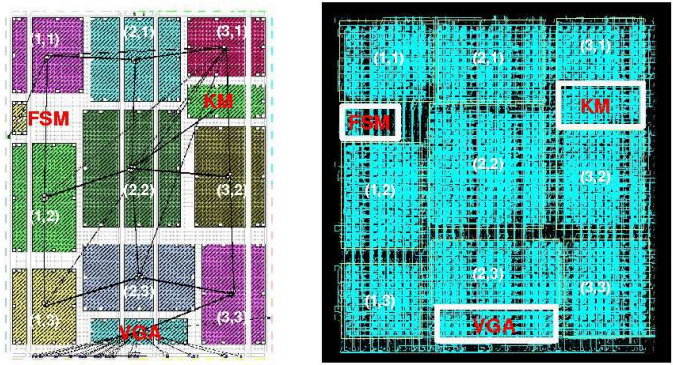,width=95mm}}
\end{minipage}
\caption{DyNoC implementation of a traffic light controller on a VirtexII-1000\label{dynoc-3x3-prot}}
\end{figure}

We have prototyped DyNoCs of different sizes and widths on two FPGAs, the
VirtexII-1000 and VirtexII-6000 from Xilinx. While the prototype on the
VirtexII-6000 seems mostly for statistically (area, latency) purpose, the
implementation on the VirtexII-1000 was done on the RC200 FPGA-board from
Celoxica. The result is given in Table \ref{table1} in terms of area (A) occupation
for different bit-widths, memory (M) usage and speed (S) in MHz.
\begin{table}[hbt]
\centering
\caption{Router Statistics}
\label{table1}
\begin{tabular}{|c|c|l|} \hline
&VirtexII-1000&VirtexII-6000\\ \hline
A/M/S(8 bit) & 8\% /4\% / 77.2&  1\% /0\% / 77.2\\ \hline
A/M/S(16 bit) & 12\% 7\% / 75.4&  2\% /1\% / 75.4\\ \hline
A/M/S(32 bit)& 21\% / 12\% / 77.3 &  3\% /2\% / 74.9\\ \hline
A/M/S(64 bit)& 46\% / 28\% / 70.1&  7\% /4\% / 73.7\\
\hline \end{tabular}
\end{table}
We have implemented two video applications with a VGA controller running at 25Mhz 
for normal 640x480 VGA. In the first one, a color generator module (CG) 
communicates with the VGA controller (VC). The color generator gets the X and Y 
coordinates of the current pixel position from the VGA module, computes
the color to be placed at that position and sends it back to the VGA module
which displays the color at the corresponding position. The color generator
application is nice for detecting changes in the communication, since this
will directly have a visual effect on the screen. The X and Y positions are coded with
12 bits each and the color with 24 bits. Therefore, we built packets with 32 bits 
width in each direction. Implemented on the RC200 board with the
DyNoC, we were not able to detect any change in the displayed pattern, event with 
a full network traffic due to the communication among remaining routers.\\
The second application is the implementation of a traffic light controller (TLC)
containing three modules: A VGA controller (VGA), a Traffic light visual module (LV) 
and a traffic control module (TC) to capture the pedestrians wishes. As in the first 
case, the VGA module is used to display the state of a traffic intersection on which 
the light and the button used by the pedestrians can be seen. The traffic visual module 
is in charge of building the traffic light infrastructure which is then displayed by the 
VGA module. The VGA sends the X and Y pixel scan positions to the traffic 
visual module and receive a color to be displayed. According to the pixel positions,
the traffic light visual computes the pattern to be placed at that position. This
generates the traffic light infrastructure. The last module is a FSM which
monitors the pedestrian inputs (two push buttons on the board) and sends a
message for the transition of state of the traffic infrastructure to the traffic light 
visual which in turn generates the corresponding color to be seen. The traffic
light controller was successfully implemented on 3x3 DyNoC. Here, we disable the router 
at position (2,2) to enforce a surrounding. All the remaining routers keep communicating 
with each other to keep the traffic high in the network. Also here, the application runs without any interruption and without any malfunction. The implementation 
of the TLC on a 3x3 DyNoC is shown in Figure \ref{dynoc-3x3-prot}.
\section{Conclusion}\label{conclusion}
In this paper we have addressed the platform of dynamic communication mechanism on on-chip
networks. A dynamic network on-chip, the DyNoC has been presented as well as a routing 
methodology able to handle obstacles in the network. The architecture and methods presented
can be used as communication medium in reconfigurable devices to solve the problem which 
arises when dynamically placed components need to communicate. We proved the feasibility 
of our approach analytically and experimentally by means of two examples. We still need to investigate the 
problem of clearing a region of the network before placing a component. This can be done by 
using global horizontal and vertical control signals to control the clearing process before 
placement.
\bibliographystyle{IEEEtran}
\bibliography{fpl2005}
\small


\end{document}